\documentclass[preprintnumbers,amsmath,11pt,amssymb,floatfix,superscriptaddress,nofootinbib]{revtex4}
\usepackage{graphicx}
\usepackage{epsfig}
\usepackage{bm}
\usepackage{amsfonts}

\input epsf

\begin{document}

\title{\Large Generalized second law of thermodynamics in the emergent universe for some viable models of $f(T)$ gravity}

\author{Rahul Ghosh}
\email{ghoshrahul3@gmail.com} \affiliation{Department of
Mathematics, Bhairab Ganguly College, Kolkata-700 056, India.}

\author{Antonio Pasqua}
\email{toto.pasqua@gmail.com} \affiliation{ Department of Physics,
University of Trieste, Via Valerio, 234127 Trieste, Italy.}

\author{Surajit Chattopadhyay}
\email{surajit_2008@yahoo.co.in, surajcha@iucaa.ernet.in}
\affiliation{Pailan College of Management and Technology, Bengal
Pailan Park, Kolkata-700 104, India.}

\date{\today}

\begin{abstract}
The present work is motivated by the study of reference
\cite{karmani}, where the generalized second law (GSL) of thermodynamics
has been investigated for a flat FRW universe for three viable
models of $f\left(T\right)$ gravity. We have here considered
a non-flat universe and, accordingly, studied the behaviors of
equation of state (EoS) parameter $\omega$ and of the deceleration parameter $q$.
Subsequently, using the first law of thermodynamics, we derived the
expressions for the time derivative of the total entropy of a
universe enveloped by apparent horizon. In the next phase, with
the choice of scale factor $a\left(t\right)$ pertaining to an emergent universe, we
have investigated the sign of the time derivatives of total
entropy for the models of $f(T)$ gravity considered.
\end{abstract}

\pacs{}

\maketitle

\section{Introduction}
The cosmic acceleration we are able to see today has been supported by many independent cosmological
observational data. The origin of Dark Energy (DE), which is widely believed to be responsible for
this cosmic acceleration, is one of the most serious problems in
modern cosmology. Nojiri and Odintsov \cite{odintsov1} reviewed
various modified gravities considered as gravitational alternative
for DE. Specifically, they \cite{odintsov1} considered
the versions of $f(R)$, $f(G)$ or $f(R, G)$ gravity models with
non-linear gravitational coupling or string-inspired model with
Gauss-Bonnet-dilaton coupling in the late universe where they lead
to cosmic speed-up. In another work, Nojiri and Odintsov
\cite{odintsov2} developed the reconstruction program for the
number of modified gravities like scalar-tensor theory, $f(R)$,
$f(G)$ and string-inspired, scalar-Gauss-Bonnet gravity. Modified
gravity with $\ln R$ or $R^{-n}(\ln R)^{m}$ terms, which grow at
small curvature, was discussed by Nojiri and Odintsov \cite{odintsov3}.
Abdalla et al \cite{odintsov4} discussed modified gravity which
includes negative and positive powers of curvature and provides
gravitational DE: he demonstrated that, in General Relativity
plus a term containing a negative power of curvature, cosmic
speed-up may be achieved. In a recently published exhaustive
review, Nojiri and Odintsov \cite{odintsov5} discussed the
structure and cosmological properties of a number of modified
theories, including traditional $f(R)$ and Horava-Lifshitz $f(R)$
gravity, scalar-tensor theory, string-inspired and Gauss-Bonnet
theory, non-local gravity, non-minimally coupled models, and
power-counting renormalizable covariant gravity. Motivated by
attempts to explain the observed acceleration of the universe in a
natural way, there has been a great deal of recent interest in a
generalization of this theory in which the Lagrangian is an
arbitrary algebraic function $f$ of the Lagrangian of teleparallel
gravity $T$ \cite{Li}. This is similar to creating $f(R)$ gravity
theories that are a generalization of general relativity (a review
on $f(R)$ gravity is available in \cite{Sotiriou}). This gravity
is dubbed as $f(T)$ gravity. Review on $f(T)$ gravity is available
in  references \cite{Zheng, Dent, Bengochea, Myrza, cai}. This
gravity describes the present accelerating expansion of the
universe without resorting to dark energy. It is a generalization
of the teleparallel gravity (TG) by replacing the so-called
torsion scalar $T$ with $f(T)$ \cite{karmani}. In
\cite{Chattopadhyay}, the EoS parameter was
reconstructed for emergent universe under $f(T)$ gravity. Bamba
and Geng \cite{bambageng} explored thermodynamics of the apparent
horizon in $f(T)$ gravity with both equilibrium and
non-equilibrium descriptions. In another work, Bamba et al
\cite{bambaetal} studied the cosmological evolutions of the
EoS for DE in the exponential and
logarithmic (as well as their combination) $f(T)$ theories. In a
recent work, Bamba et al \cite{bambaetal1} reconstructed a model
of $f(T)$ gravity with realizing the finite-time future
singularities. In addition, they \cite{bambaetal1} have explicitly
shown that a power-law-type correction term $T^{\beta}$ (with
$\beta>1$) such as a $T^{2}$ term can remove the finite-time
future singularities in $f(T)$ gravity.\\
The possibilities of an emergent universe
\cite{Ellis1,Ellis2,Paul} have been studied recently in a number
of papers in which one looks for an ever-existing
and large enough universe so that the space-time may be treated as classical
entities. In these models, the universe in the infinite past is in
an almost static state but it eventually evolves into an
inflationary stage \cite{Paul}. An emergent universe model can be
defined as a singularity free universe which is ever existing with
an almost static nature in the infinite past
$(t\rightarrow-\infty)$ and then evolves into an inflationary
stage \cite{Debnath}. In references \cite{Paul, Canadian,
Chattopadhyay} the characteristics of an emergent universe have
been summarized as:
\begin{enumerate}
    \item the universe is almost static at the finite past $(t\rightarrow-\infty)$ and isotropic and homogeneous at large
    scales;
    \item it is ever existing and there is no timelike
    singularity;
    \item the universe is always large enough so that the classical
description of space-time is adequate;
    \item the universe may contain exotic matter so that the energy conditions
may be violated;
    \item the universe is accelerating as suggested by recent measurements of distances of
high redshift type Ia supernovae.
\end{enumerate}
Chattopadhyay and Debnath \cite{Canadian} considered the generalized Ricci DE and generalized holographic DE (HDE) in the scenario of an emergent universe and they studied the behaviors of the potential
and the chameleon scalar fields. In references \cite{Debnath,
Canadian, Chattopadhyay}, the choice of scale factor $a\left(t\right)$ for
emergent universe was $a\left(t\right)=a_{0}\left(\eta+e^{B t}\right)^{n}$,
where $a_{0}$, $\eta$, $B$ and $n$ are positive constants.

    Brevik et al \cite{odintsov5} studied the entropy of a FRW universe
filled with dark energy (cosmological constant, quintessence, or
phantom). Bamba et al \cite{bambaodintsov} discussed the relation
between the expression of the entropy and the contribution from
the modified gravity as well as the matter to the definition of
the energy flux (heat).

    The present work is motivated by the work of \cite{karmani}, who
investigated the validity of the generalized second law (GSL) of
gravitational thermodynamics in the framework of $f(T)$ gravity.
The present study deviates the earlier work in the following
aspects: (i) we have considered three viable models of $f(T)$
gravity without the assumption of flat universe and examined the
cases of non-flat universe; (ii) we have examined the validity of
GSL of thermodynamics for a particular choice of scale factor
pertaining to emergent universe; (iii) we have investigated some
particular cases for the violation of GSL.

\section{$f(T)$ gravity in non-flat universe}
In the framework of $f(T)$ theory, the action $I$ of modified TG is
given by \cite{karmani}:
\begin{equation}
I=\frac{1}{16\pi G}\int d^{4}x \sqrt{-g}\left[f(T)+L_{m}\right]
\end{equation}
where $L_{m}$ is the Lagrangian density of the matter inside the
universe.
We consider a Friedmann-Robertson-Walker (FRW) universe
filled with the pressureless matter (i.e., $p_m=0$). Choosing $(8\pi G=1)$ the
modified Friedman equations in the framework of $f(T)$ gravity are
given by \cite{karmani, Ferraro}:
\begin{eqnarray}
H^{2}+\frac{k}{a^{2}} &=& \frac{1}{3}\left(\rho_{m}+\rho_{T}\right), \\
\dot{H}-\frac{k}{a^{2}} &=& -\frac{1}{2}\left(\rho_{m}+\rho_{T}+p_{T}\right),
\end{eqnarray}
where:
\begin{eqnarray}
\rho_{T} &=& \frac{1}{2}(2T f_{T}-f-T), \\
p_{T} &=& -\frac{1}{2}\left[-8\dot{H}T f_{TT}+(2T-4\dot{H})f_{T}-f+4\dot{H}-T\right].
\end{eqnarray}
The torsion scalar $T$ is defined, for non-flat universe, as \cite{Ferraro}:
\begin{equation}
T=-6\left(H^{2}+\frac{k}{a^{2}}\right),
\end{equation}
In above Equations, $k$ is the curvature parameter, which can assume the values $\left(-1, 0, +1\right)$,
$H=\frac{\dot{a}}{a}$ is the Hubble parameter, $\dot{H}$ is the derivative with respect to the cosmic time $t$
of the Hubble parameter $H$, $f_T$ is the first derivatie of $f$ with respect to $T$,
$f_{TT}$ if the second derivative of $f$ with respect to $T$ and $\rho_{m}$ is
the energy density of the matter. Moreover, $\rho_{T}$ and $p_{T}$ are
the torsion contributions to the energy density and pressure.
Energy conservation equations are given by:
\begin{equation}
\dot{\rho}_{m}+3H\rho_{m}=0, \nonumber \\
\dot{\rho}_{T}+3H(\rho_{T}+p_{T})=0.
\end{equation}
The effective EoS parameter due to the torsion
contribution is defined as \cite{karmani}:
\begin{equation}
\omega_{T}=-1-\frac{\dot{T}}{3H}\left(\frac{2T
f_{TT}+f_{T}-1}{2Tf_{T}-f-T}\right).
\end{equation}
From Eqs.  (2), (4) and (6) we get $\rho_{m}$ as:
\begin{equation}
\rho_{m}=\frac{1}{2}\left(f-2Tf_{T}\right).
\end{equation}
Using Eqs. (3) and (9), we derive that the time derivative of Hubble parameter $H$ is given by:
\begin{equation}
\dot{H}=-\frac{1}{2}\left(\frac{\rho_{m}-\frac{2k}{a^{2}}}{f_{T}+2Tf_{TT}}\right).
\end{equation}
Using Eqs. (6) and (10), we derive the following expression for $\dot{T}$:
\begin{equation}
\dot{T}=\frac{12
H}{f_{T}+2Tf_{TT}}\left[\frac{k(f_{T}+2Tf_{TT}-1)}{a^{2}}+\frac{\rho_{m}}{2}\right].
\end{equation}
The deceleration parameter $q$ can be obtained from  Eqs. (2), (9) and (10)  as follow:
\begin{equation}
q=-1-\frac{\dot{H}}{H^{2}}=-1+\frac{3}{2}\left\{\frac{\rho_{m}a^{2}-2k}{\left(f_{T}+2Tf_{TT}\right)\left[a^{2}\left(\rho_{m}+\rho_{T}\right)-3k\right]}\right\}.
\end{equation}
\section{GSL in $f(T)$ gravity}
The apparent horizon has been argued as a causal horizon for a
dynamical spacetime and it is associated with gravitational entropy
and surface gravity \cite{Bak}. The radius of the apparent
horizon, which is denoted by $R_{A}$, is given by \cite{sheykhi}:
\begin{equation}
R_{A}=\frac{1}{\sqrt{H^{2}+\frac{k}{a^{2}}}}.
\end{equation}
In the present work, we are investigating the validity of GSL in a
non-flat FRW universe filled with pressureless Dark Matter (DM). The
GSL can be expressed as $\dot{S}_{A}+\dot{S}_{m}\geq 0$, where
$S_{A}$ denotes the Bekenstein-Hawking entropy on the horizon and
$S_{m}$ is the entropy due to the matter sources inside the
horizon \cite{akbar}. Detailed account of the GSL is available in
\cite{sheykhi, jamil, paddy}. From the first law of thermodynamics
one can get (Clausius relation) $-dE = T_{A}dS_{A}$ to the
apparent horizon $R_{A}$. The Friedmann equation in the Einstein
gravity can be derived if we take the Hawking temperature $T_{A}
=1/2\pi R_{A}$ and the entropy $S_{A}=A/4G$ on the apparent
horizon, where $A$ is the area of the horizon \cite{karmani}.
However, this definition is changed for other modified gravity
theories. In $f(T)$ gravity, it was shown that when $f_{TT}$ is
small, the first law of black hole thermodynamics is satisfied
approximatively and the entropy of horizon is
$S_{A}=\frac{Af_{T}}{4G}$ \cite{karmani}. In the
present work we assume that the boundary of the universe to be
enclosed by the dynamical apparent horizon $R_{A}$ and the Hawking
temperature on the apparent horizon is given by \cite{karmani,
sheykhi}:
\begin{equation}
T_{A}=\frac{1}{2\pi
R_{A}}\left(1-\frac{\dot{R}_{A}}{2HR_{A}}\right).
\end{equation}
Based on the discussions in the previous Section, we compute:
\begin{eqnarray}
T_{A}\dot{S}_{m} &=& 2\pi R_{A}^{2}(f-2Tf_{T})(\dot{R}_{A}-HR_{A}), \\
T_{A}\dot{S}_{A}&=&4\pi
\left(1-\frac{\dot{R}_{A}}{2HR_{A}}\right)\left[2\dot{R}_{A}f_{T}+6H\left(\frac{\rho_{m}-\frac{2k}{a^{2}}}{f_{T}+2Tf_{TT}}\right)R_{A}f_{TT}\right],
\end{eqnarray}
where $\dot{S}_{m}$ and $\dot{S}_{A}$ denote, respectively, the time derivatives
of the entropy due to the matter sources inside the horizon and on
the horizon. In the following subsections, we investigate
whether $\dot{S}_{m}+\dot{S}_{A}\geq 0$ for three viable choices
of $f(T)$ models with the choice of scale factor for emergent scenario,
i.e. $a=a_{0}\left(\eta+e^{B t}\right)^{n}$. The $f(T)$ models considered here are:
\begin{enumerate}
    \item $f(T)=\gamma T+\lambda T^{m}$, where $\gamma$, $\lambda$ and $m$ are constants \cite{Chattopadhyay}
    \item $f(T)=T-\lambda T \left[1-e^{\left(\beta\frac{T_{0}}{T}\right)}\right]$ where
    $\lambda=\frac{1-8\pi
    \frac{\rho_{m0}}{3H_{0}^{2}}}{1-(1-2\beta)e^{\beta}}$\cite{karmani}
    \item $f(T)=T+\lambda(-T)^{n_{1}}$ where
   $\lambda=\left(\frac{1-8\pi
   \frac{\rho_{m0}}{3H_{0}^{2}}}{2n_{1}-1}\right)(6H_{0}^{2})^{(1-n_{1})}$ and $n_1$ is a constant \cite{karmani}
\end{enumerate}
$T_0$, $\rho_{m0}$ and $H_0$ are, respectively, the present values of $T$, energy density of DM and Hubble parameter.

\subsection{Model I}
First, we consider the model $f(T)=\gamma T+\lambda T^{m}$. In this case, we
have that $\dot{T}$ is given by:
\begin{equation}
  \dot{T}=\frac{\xi_{1}}{\xi_{2}},
\end{equation}
where:
\begin{equation}
\begin{array}{c}
 \xi_{1}=6Be^{Bt}n(e^{Bt}+\eta)^{-3-2n}\left[6(k^{2}(2+\gamma)(e^{Bt}+\eta)^{4}+3a_{0}^{4}B^{4}e^{4Bt}n^{4}\gamma\times\right.\\
\left.(e^{Bt}+\eta)^{4n}-2a_{0}^{2}B^{2}e^{2Bt}kn^{2}(1+2\gamma)(e^{Bt}+\eta)^{2+2n})+a_{0}^{2}(e^{Bt}+\eta)^{2n}\times\right.\\
\left.\left\{-\frac{B^{2}e^{2Bt}n^{2}}{(e^{Bt}+\eta)^{2}}+\frac{k(e^{Bt}+\eta)^{-2n}}{a_{0}^{2}}\right\}^{m}(-6^{m}k(3+4(-2+m)m)(e^{Bt}+\eta)^{4}+3a_{0}^{2}B^{2}e^{2Bt})\lambda \right],\\
\end{array}
\end{equation}

\begin{equation}
\begin{array}{c}
  \xi_{2}=a_{0}^{2}\left[-6k\gamma(e^{Bt}+\eta)^{2}+6a_{0}^{2}B^{2}e^{2Bt}n^{2}\gamma(e^{Bt}+\eta)^{2n}+a_{0}^{2}m(6^{m}-2^{1+m}3^{m}m)(e^{Bt}+\eta)^{2+2n}\lambda\times\right.\\
\left.\left\{-\frac{B^{2}e^{2Bt}n^{2}}{(e^{Bt}+\eta)^{2}}+\frac{k(e^{Bt}+\eta)^{-2n}}{a_{0}^{2}}\right\}^{m}\right].\\
\end{array}
\end{equation}

\begin{figure}[h]
\begin{minipage}{16pc}
\includegraphics[width=16pc]{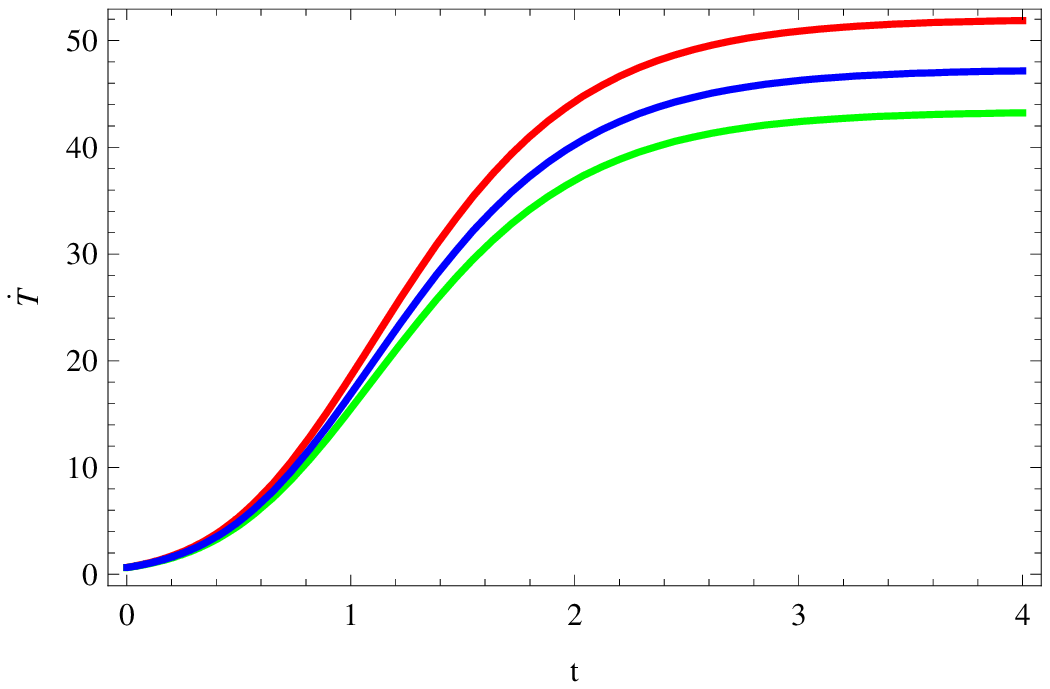}
\caption{\label{label}Evolution of $\dot{T}$ with cosmic time $t$
for model I.}
\end{minipage}\hspace{2pc}%
\begin{minipage}{16pc}
\includegraphics[width=16pc]{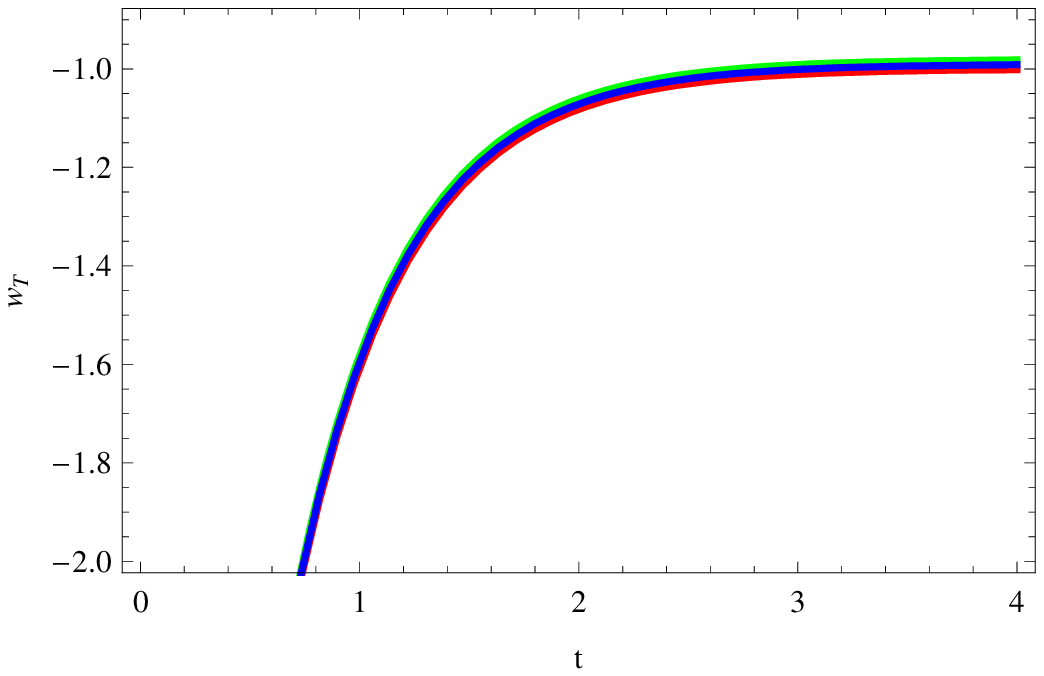}
\caption{\label{label}Behavior of $\omega_{T}$ for model I.}
\end{minipage}\hspace{3pc}%
\end{figure}

\begin{figure}[h]
\begin{minipage}{16pc}
\includegraphics[width=16pc]{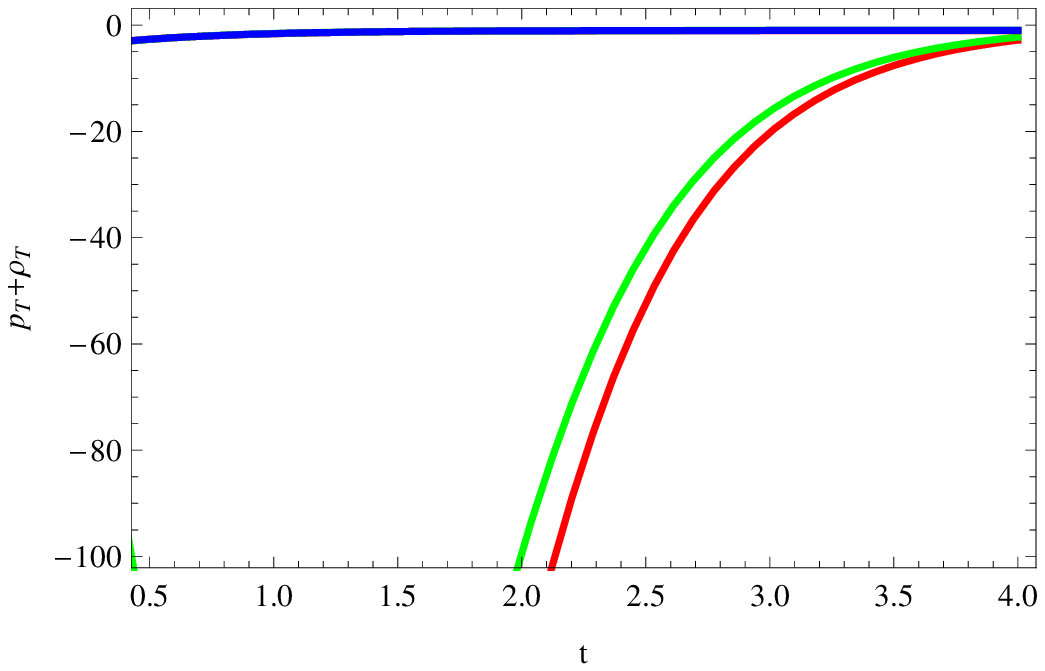}
\caption{\label{label}Behavior of $p_{T}+\rho_{T}$ for model I.}
\end{minipage}\hspace{2pc}%
\begin{minipage}{16pc}
\includegraphics[width=16pc]{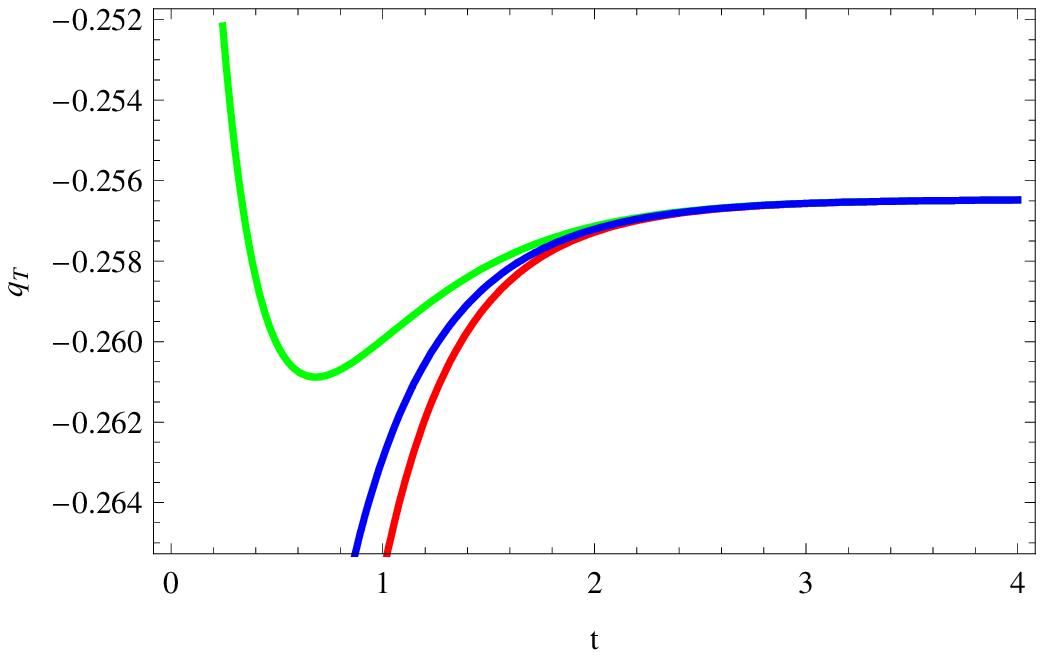}
\caption{\label{label}Behavior of $q_{T}$ for model I.}
\end{minipage}\hspace{3pc}%
\end{figure}

\begin{figure}
\includegraphics[height=2.0in]{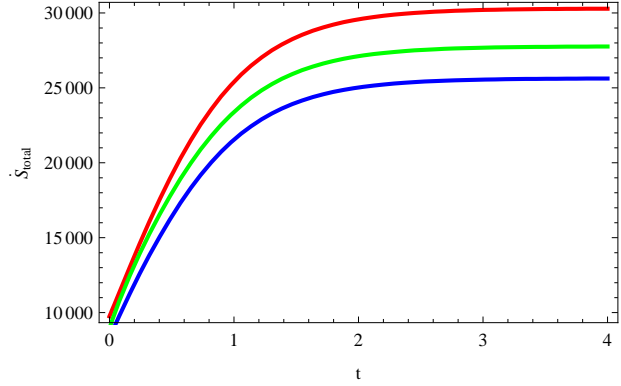}\\
\caption{Behavior of $\dot{S}_{total}$ when the apparent horizon
is considered as the enveloping horizon under $f(T)=\gamma
T+\lambda T^{m}$ i.e. model I.}
\end{figure}

In Figure 1, we plotted the behavior of
$\dot{T}$ for model I and it
 is found that $\dot{T}$ is  an increasing function of the time $t$.  The EoS parameter
 for this model has been plotted in Figure 2, where it is observed
 that $\omega_{T}\leq -1$, which indicates phantom-like
 behavior. Moveover, it is observed that from early to late stage of
 the universe the EoS parameter is tending towards $-1$.
 This holds for $k=-1,~+1$ and $0$. The deceleration parameter $q$,
 plotted in Figure 3, remains negative throughout the evolution of
 the universe. This indicates ever-accelerating
 universe. For all of the above Figures, we have considered the following values for the involved parameters:
 $B=2,~a_{0}=1.5,~n=1.2,~\eta=3,~\gamma=4.02,~\lambda=4.03$ and
 $m=2$. The sum of the pressure $p_T$ and the energy density $\rho_T$  is plotted
 in Figure 4, where we can see that $\rho_T + p_T$ is always negative, indicating
 the violation of the strong energy condition. In Figure 5, we plotted the time derivative of total
 entropy $\dot{S}_{total}$ against cosmic time $t$. $\dot{S}_{total}$ is computed using Eqs. (15) and (16), where $f(T)$ has been replaced
 by the form of model I. $\dot{S}_{total}$ is found to be positive
 throughout the evolution of the universe, which indicates that the
 GSL is satisfied by model I under emergent scenario of the
 universe.
\subsection{Model II}
In this Section, we consider the model II, given by $f(T)=T-\lambda T
\left[1-e^{\left(\beta\frac{T_{0}}{T}\right)}\right]$. The expression of $\dot{T}$ for this model is not reported here since it would occupy too many space.  $\dot{T}$ is
plotted against cosmic time $t$ in Figure 1: it is observed a behavior similar
to that for model I. It may be noted that for model II,
our choices for the parameters are $a_{0}=20,~B=4,~n=0.04$ and $\eta=3$ that satisfy
the conditions for emergent universe. Moreover, we choose
$\lambda=0.03,~\beta=-0.02,~\rho_{m0}=0.23$ and $H_{0}=74.2$. In
all Figures for this model, the red, green and blue lines correspond to the
cases $k=-1$, $k=+1$ and $k=0$, respectively.
\begin{figure}[h]
\begin{minipage}{16pc}
\includegraphics[width=16pc]{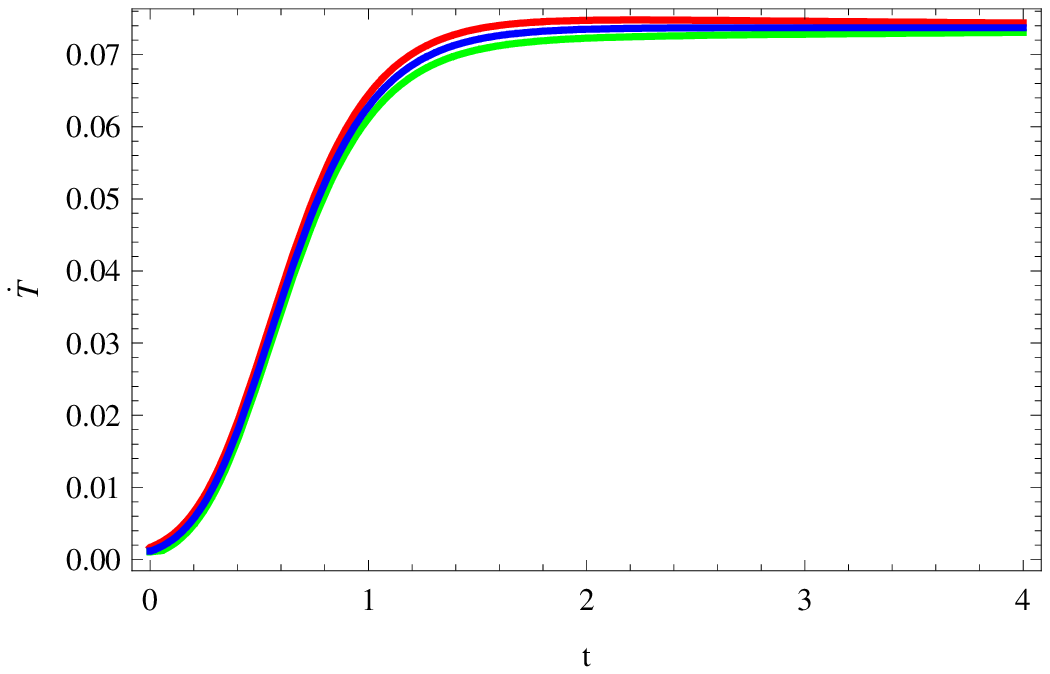}
\caption{\label{label}Behavior of $\dot{T}$ for model II.}
\end{minipage}\hspace{2pc}%
\begin{minipage}{16pc}
\includegraphics[width=16pc]{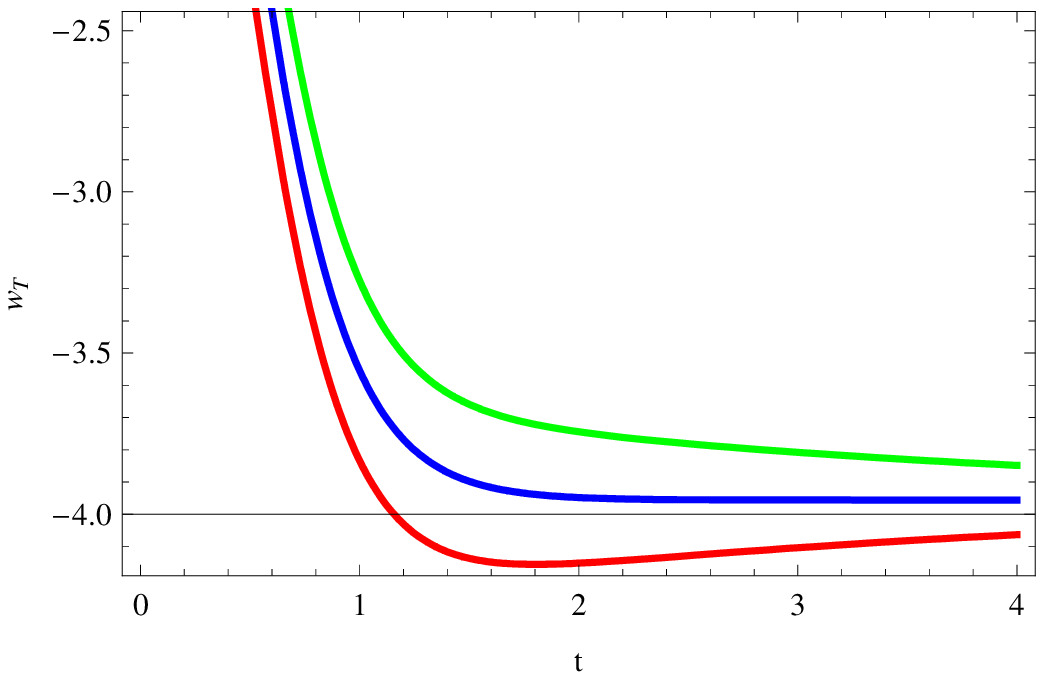}
\caption{\label{label}Behavior of $\omega_{T}$ for model II.}
\end{minipage}\hspace{3pc}%
\end{figure}
Like model I, $\dot{T}$ (plotted in Figure 6) stays positive and exhibits
upward movement with evolution of the universe. The EoS parameter $\omega_T$,
as seen in Figure 7, stays below $-1$ and it never tend to $-1$.
This indicates the phantom-like behavior of $\omega_T$.
\begin{figure}[h]
\begin{minipage}{16pc}
\includegraphics[width=16pc]{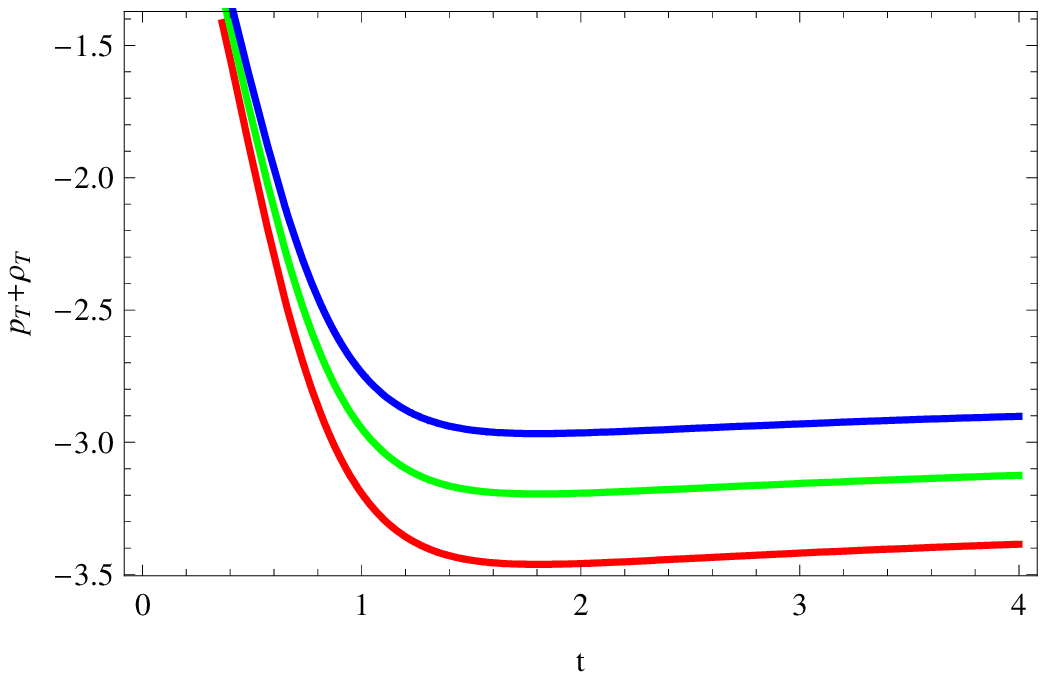}
\caption{\label{label}Behavior of $(p_{T}+\rho_{T})$ for model
II.}
\end{minipage}\hspace{2pc}%
\begin{minipage}{16pc}
\includegraphics[width=16pc]{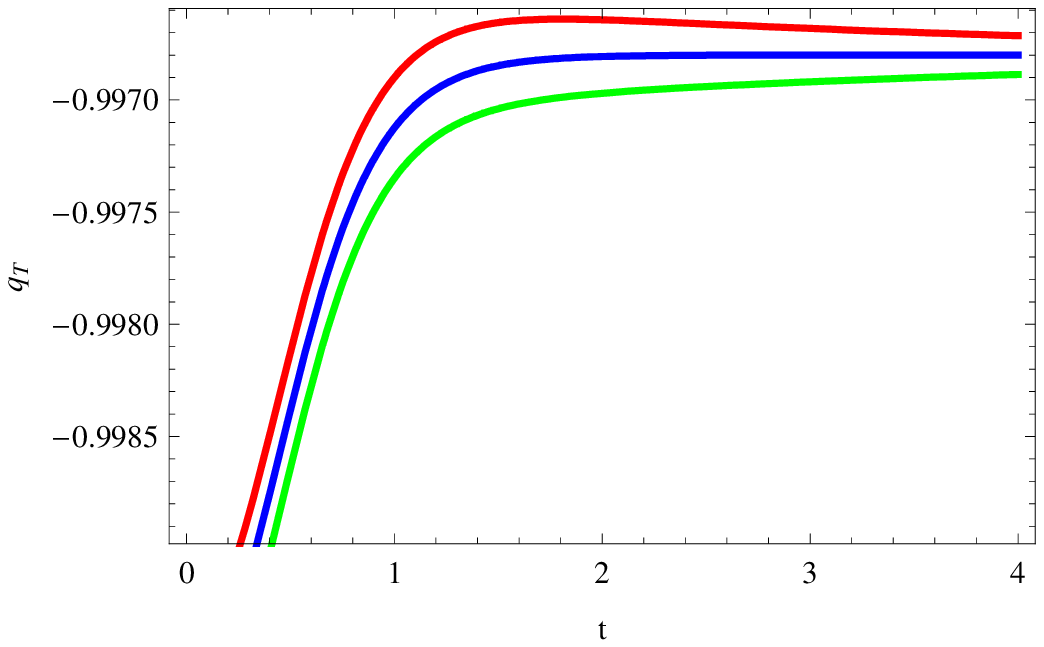}
\caption{\label{label}Behavior of $q_{T}$ for model II.}
\end{minipage}\hspace{3pc}%
\end{figure}
In Figure 8, $(p_{T}+\rho_{T})$ is plotted against $t$: it shows
the violation of strong energy condition by model II. The
accelerated universe under model II is understandable from Figure
9 which shows that $q_{T}$ is always negative for every choice of the curvature parameter $k$.
\begin{figure}
\includegraphics[height=2.0in]{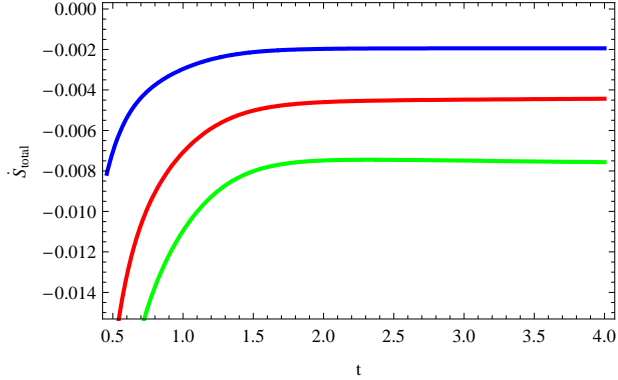}\\
\caption{Behavior of $\dot{S}_{total}$ when the apparent horizon
is considered as the enveloping horizon under $f(T)=T-\lambda T
\left[1-e^{\left(\beta\frac{T_{0}}{T}\right)}\right]$, i.e. model
II.}
\end{figure}
Finally, when we consider the time derivative of total entropy
$\dot{S}$ in Figure 10, we find that it remains negative for every
choice of $k$. This indicates violation of GSL by model II.
\subsection{Model III}
In this Section we consider model III where
$f(T)=T+\lambda(-T)^{n_{1}}$, where $\lambda=\left(\frac{1-8\pi
\frac{\rho_{m0}}{3H_{0}^{2}}}{2n_{1}-1}\right)(6H_{0}^{2})^{(1-n_{1})}$.
In order to study the model III, we have taken both $n_{1}<1~
(n_{1}=\frac{1}{3})$ and $n_{1}>1~(n_{1}=1.2)$ into account. The expression of $\dot{T}$ for this model is not reported here since it would occupy too many space.
We have plotted $\dot{T}$ in Figure 11, where
$\dot{T}$ is found to behave similarly to the cases corresponding to the
models I and II.
\begin{figure}
\includegraphics[height=2.0in]{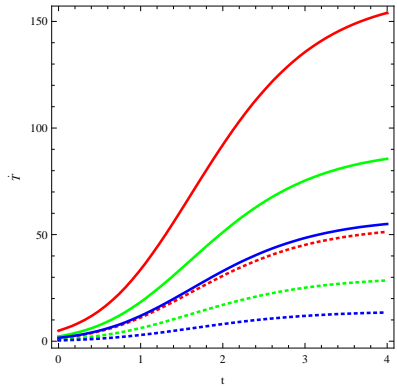}\\
\caption{Behavior of $\dot{T}$ for model III i.e.
$f(T)=T+\lambda(-T)^{n_{1}}$. The solid lines pertain to $n_{1}<1$
and the dashed lines pertain to $n_{1}>1$.}
\end{figure}
From Figures 12 and 13, we find that the EoS parameter $\omega_{T}$ has a
phantom-like behavior and it tends to $-1$ in both cases, irrespective of
the choice of the curvature parameter $k$.
\begin{figure}[h]
\begin{minipage}{16pc}
\includegraphics[width=16pc]{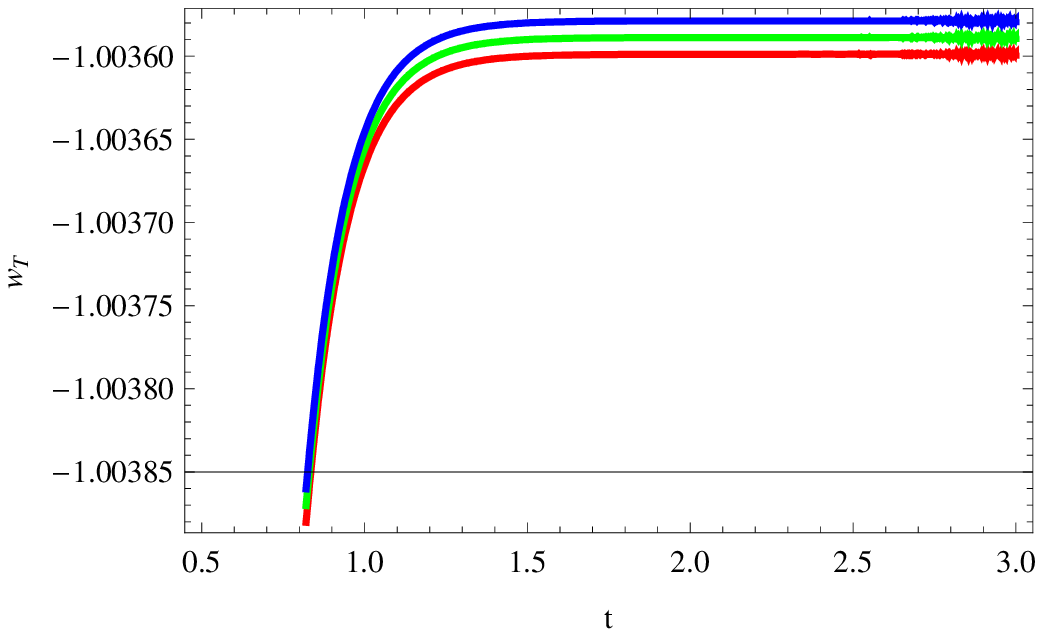}
\caption{\label{label}Behavior of $\omega_{T}$ for model III with
$n_{1}<1$.}
\end{minipage}\hspace{2pc}%
\begin{minipage}{16pc}
\includegraphics[width=16pc]{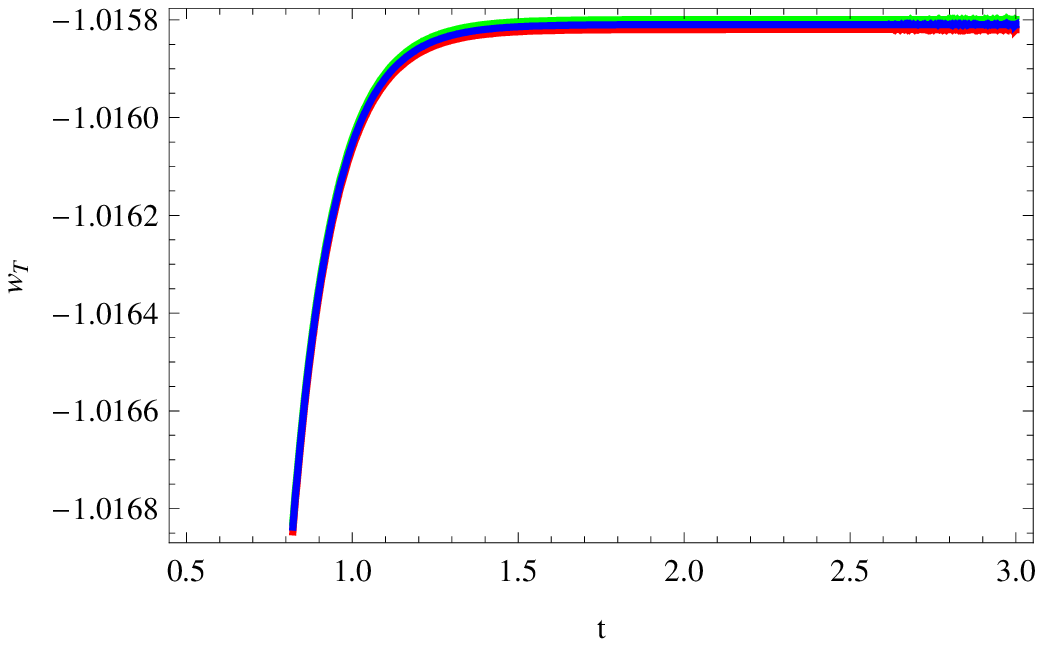}
\caption{\label{label}Behavior of $\omega_{T}$ for model III with
$n_{1}>1$.}
\end{minipage}\hspace{3pc}%
\end{figure}
The deceleration parameter $q$, plotted in Figures 14 and 15 for the two different values of $n_1$, indicates
 the realization of the accelerated expansion of the universe for
model III forboth  $n_{1}<1$ and $n_{1}>1$.
\begin{figure}[h]
\begin{minipage}{16pc}
\includegraphics[width=16pc]{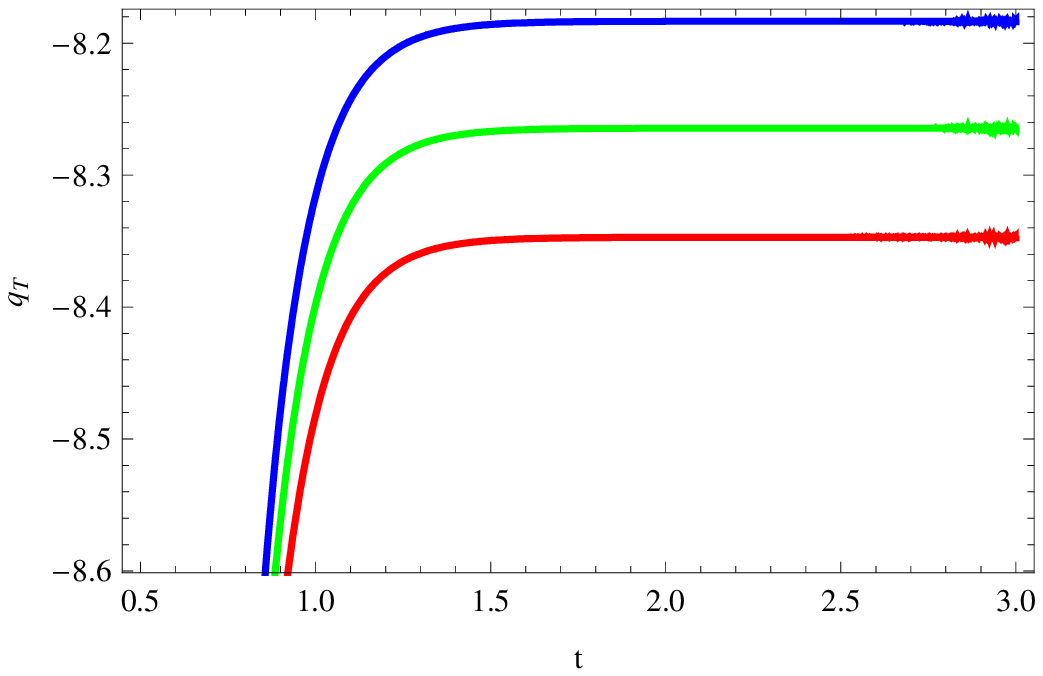}
\caption{\label{label}Behavior of $q_{T}$ for model III with
$n_{1}<1$.}
\end{minipage}\hspace{2pc}%
\begin{minipage}{16pc}
\includegraphics[width=16pc]{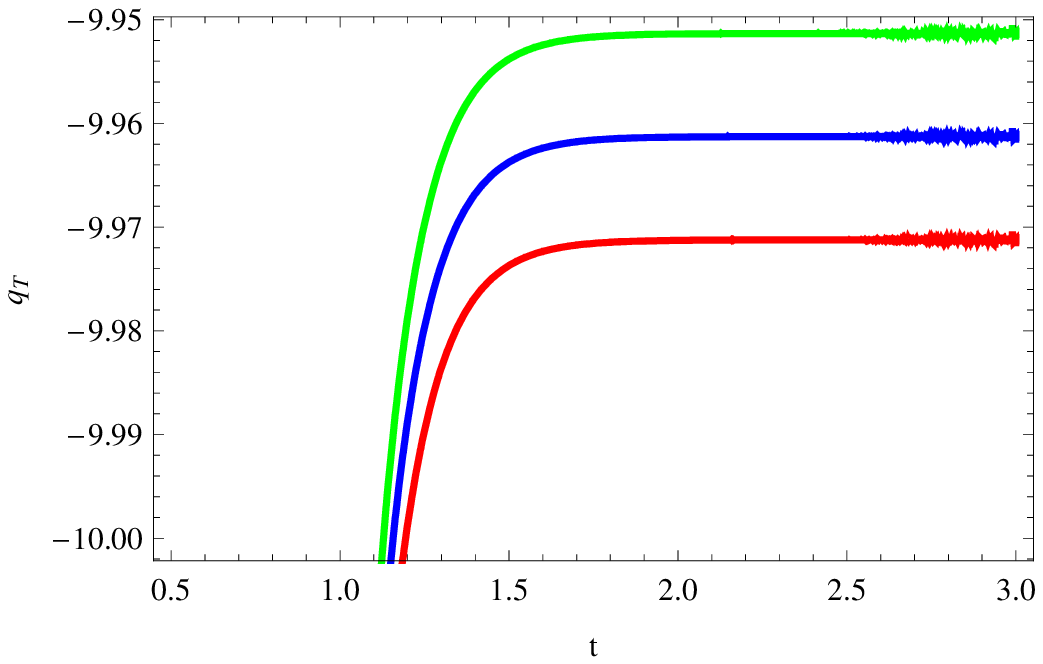}
\caption{\label{label}Behavior of $q_{T}$ for model III with
$n_{1}>1$.}
\end{minipage}\hspace{3pc}%
\end{figure}

\begin{figure}[h]
\begin{minipage}{16pc}
\includegraphics[width=16pc]{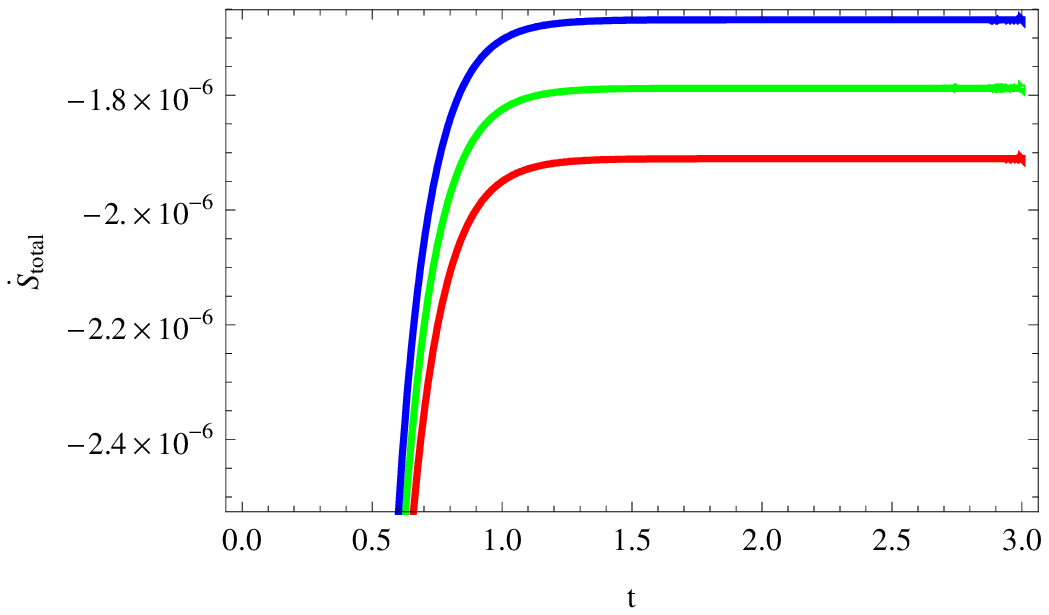}
\caption{\label{label}Behavior of $\dot{S}_{total}$ when the
apparent horizon is considered as the enveloping horizon under
$f(T)=T+\lambda(-T)^{n_{1}}$ with $n_{1}<1$.}
\end{minipage}\hspace{2pc}%
\begin{minipage}{16pc}
\includegraphics[width=16pc]{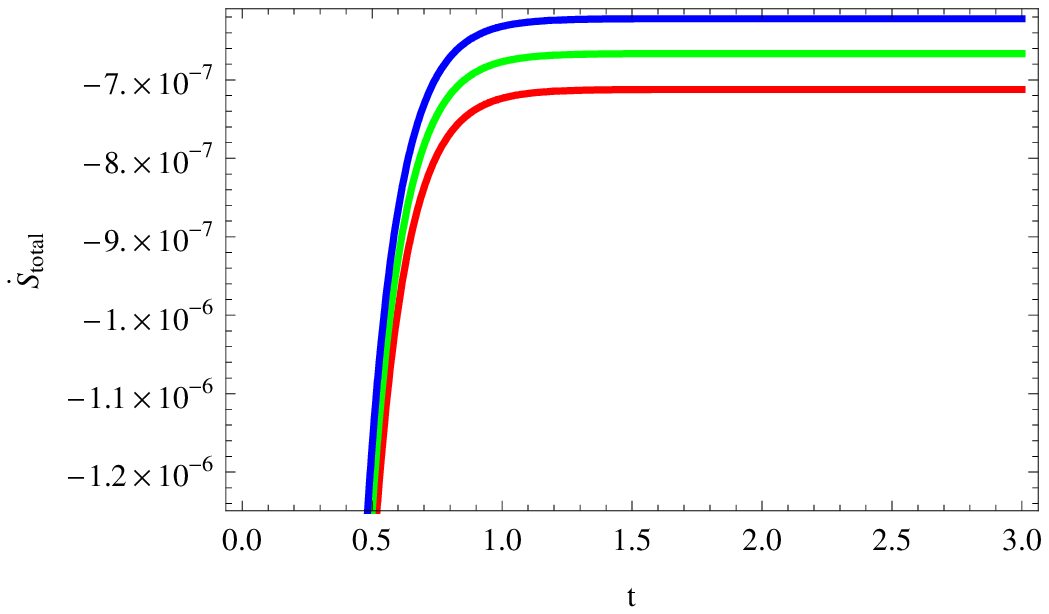}
\caption{\label{label}Behavior of $\dot{S}_{total}$ when the
apparent horizon is considered as the enveloping horizon under
$f(T)=T+\lambda(-T)^{n_{1}}$ with $n_{1}>1$.}
\end{minipage}\hspace{3pc}%
\end{figure}
Finally, when we plot $\dot{S}_{total}$ in Figures 16 and 17 for
the case of model III with apparent horizon as the enveloping
horizon, we observe violation of the GSL by model III irrespective
of the choices of $n_{1}$.

\section{Discussion}
In this work, we have considered three viable $f(T)$ gravity with
scale factor $a$ pertaining to emergent universe. We have
investigated the validity of GSL for all the cases along with some
related characteristics. It is observed that for model I, $\dot{T}$ exhibits an increasing pattern in the positive
side with evolution of the universe. When we considered the EoS
parameter $\omega_T$ for this model, we found that, with evolution of the
universe, $\omega_{T}$ remains below -1 at earlier stages and tends
to -1 at later stage, indicating a phantom-like behavior of
$\omega_{T}$. As displayed in Figure 2. In Figure 3 we plotted
$\left(p_{T}+\rho_{T}\right)$ against the cosmic time $t$ and we found that it
stays at negative level. This indicates violation of strong
energy condition. The deceleration parameter $q_{T}$ for this
model is plotted against cosmic time $t$ in Figure 4: we found
that it is negative throughout. This indicates accelerated phase
of the universe. The time derivative of total entropy is plotted
in Figure 5 and we find that it remains positive through the
evolution of the universe irrespective of the value of $k$.
Therefore our observations are that for model I are: (i) the EoS
exhibits phantom-like behavior, (ii) the strong energy condition
is violated, (iii) we get an accelerating universe and (iv)
the GSL of thermodynamics holds irrespective of the
curvature of the universe. It should be stated that the apparent
horizon has been considered as the enveloping horizon of the
universe. Now we consider the model II with the choices of scale
factor and enveloping horizon similar to that of model I. All the
quantities considered for model I are also considered for model II. From Figure 6, we observe that the behavior of $\dot{T}$
is similar to that of model I. However
the EoS parameter $\omega_T$, as plotted in Figure 7, exhibits a small difference
with model I. In the present case, the EoS parameter $\omega_T$ exhibits
phantom-like behavior. However, it tends to -1 at later stages
of the universe. The strong energy condition is violated like in
model I (Figure 8). Moreover, like in model I, we also find an ever
accelerating universe (see Figure 9). Finally, when we consider the
GSL, we find a prominent difference between model I and model II.
In model II, the time derivative of total entropy remains negative
through the evolution of the universe. This indicates the
violation the GSL. This is displayed in Figure 10. Therefore for
model II the observations are following: (i) the EoS exhibits
phantom-like behavior, (ii) The strong energy condition is
violated, (iii) we get accelerating universe and (iv) generalized
second law of thermodynamics is violated here. Here also the above
results hold irrespective of the choice of $k$. Now we come to
Model III. For model III we have same choices of scale factor and
enveloping horizon. However, while considering model III we
consider two cases, namely $n_{1} <1$ and $n_{1} >1$. From Figure
11 we find that the time derivative of $T$ remains positive for
$n_{1} <1$ as well as $n_{1} >1$ and, in both cases, it shows
increasing pattern. In Figure 12 we have plotted EoS parameter for
$n_{1}<1$ and we have observed that it is always staying below
$-1$ irrespective of the choice of curvature. This indicates
phantom-like behavior of the EoS parameter for $n_{1} <1$. In
Figure 13, we plot the EoS parameter for $n_{1} >1$ and here also we observe
phantom like behavior. In Figure 14 and 15 we plot the
deceleration parameter $q_{T}$ for $n_{1} <1$ and $n_{1} >1$: in both cases, we get a negative deceleration
parameter $q$, indicating ever accelerating universe. The GSL for $n_{1} <1$ is considered in Figure 16 and we
observe that the time derivative of total entropy is staying at
negative level, which means a violation of the GSL. The case corresponding to $n_{1}
>1$ is considered in Figure 17. Here, we also find that the
time derivative of the total entropy is negative through the
universe. Therefore, we conclude that the GSL is not valid in this
case. The above results hold for $k=-1,~+1$ and $0$. Therefore,
our observations are the following: (i) the EoS parameter has a
phantom-like behavior, (ii) the accelerated expansion of the universe is realized
and (iii) the generalized second law of thermodynamics does not
work here. The above results are independent of the choice of the curvature parameter $k$.
\\\\
\section{Acknowledgements}
The first author wishes to thank the Inter-University Centre for
Astronomy and Astrophysics (IUCAA), Pune, India for providing warm
hospitality during a scientific visit in January 2012, when part
of the work was carried out. The third author sincerely
acknowledges the Visiting Associateship provided by IUCAA, Pune,
India for the period of August 2011 to July 2014 to carry out
research in General Relativity and Cosmology. The third author
acknowledges the research grant under Fast Track Programme for
Young Scientists provided by the Department of Science and
Technology (DST), Govt of India. The project number is
SR/FTP/PS-167/2011.
\\\\

\end{document}